\documentclass[10pt,aps,pra,twocolumn,superscriptaddress]{revtex4-1}
\usepackage{amsmath}
\usepackage{amssymb}
\usepackage{graphicx}
\usepackage{hyperref}
\hypersetup{pdfstartview={FitH},pdfpagemode={UseNone},
            colorlinks,linkcolor=blue, citecolor=blue, urlcolor=blue,
             bookmarksopen=true, pdfnewwindow=true}

\graphicspath{{Figures/}{}}

\usepackage{braket}

\begin{document}

\title{Reconfigurable cluster state generation in specially poled nonlinear waveguide arrays }

\author{James G. Titchener}
\affiliation{Nonlinear Physics Centre, Research School of Physics and Engineering, Australian National University, Canberra, ACT 2601, Australia}
\affiliation{Quantum Technology Enterprise Centre, Quantum Engineering Technology Labs, H. H. Wills Physics Laboratory and Department of Electrical and Electronic Engineering, University of Bristol, BS8 1FD, UK}

\author{Alexander S. Solntsev}
\affiliation{Nonlinear Physics Centre, Research School of Physics and Engineering, Australian National University, Canberra, ACT 2601, Australia}
\affiliation{School of Mathematical and Physical Sciences, University of Technology Sydney, 15 Broadway, Ultimo NSW 2007, Australia}

\author{Andrey A. Sukhorukov}
\affiliation{Nonlinear Physics Centre, Research School of Physics and Engineering, Australian National University, Canberra, ACT 2601, Australia}

%\affiliation{Corresponding author: James.Titchener@anu.edu.au}

%\dates{Compiled \today}

%\ociscodes{(190.4390)   Nonlinear optics, integrated optics; (270.5585)   Quantum information and processing.}

%\doi{\url{http://dx.doi.org/10.1364/ao.XX.XXXXXX}}

\begin{abstract}
We present a new approach for generating cluster states on-chip, with the state encoded in the spatial component of the photonic wavefunction. We show that for spatial encoding, a change of measurement basis can improve the practicality of cluster state algorithm implementation, and demonstrate this by simulating Grover's search algorithm.
Our state generation scheme involves shaping the wavefunction produced by spontaneous parametric down-conversion in on-chip waveguides using specially tailored nonlinear poling patterns.
Furthermore the form of the cluster state can be reconfigured quickly by driving different waveguides in the array. Importantly, this approach allows cluster states to be generated directly from a nonlinear optical process, without requiring additional optical transformations to be applied after the initial quantum state is generated.

\end{abstract}

\maketitle

%\setboolean{displaycopyright}{true}

%\begin{document}

\maketitle
%\thispagestyle{fancy}

%\ifthenelse{\boolean{shortarticle}}{\ifthenelse{\boolean{singlecolumn}}{\abscontentformatted}{\abscontent}}{}

\section{INTRODUCTION}

Cluster states are highly entangled multi-particle quantum states~\cite{Briegel:2001-910:PRL} that have drawn significant interest for their potential in quantum information processing~\cite{Raussendorf:2001-5188:PRL,Raussendorf:2003-22312:PRA}. These multi-qubit states form a complete basis for one way quantum computation, where algorithms are carried out by successive measurement of qubits, causing information to flow through the state via entanglement~\cite{Nielsen:2006-147:RMAP}. Crucially for practical applications, cluster states have been shown to be robust to decoherence and loss of qubits~\cite{Hein:2005-32350:PRA}. In solid state physics cluster states are naturally produced in spin lattices interacting by an Ising type Hamiltonian~\cite{Briegel:2001-910:PRL,Mandel:2003-937:NAT,Lanyon:2013-210501:PRL}, but increasingly they are considered useful in quantum photonic systems \cite{Gimeno-Segovia:2015-20502:PRL}.

In quantum photonics the nonlinear interactions required for gates between single photons are challenging to realize~\cite{Feizpour:2015-905:NPHYS}, while measurement based approaches to quantum computation, including those employing cluster states, can be more readily implemented~\cite{Nielsen:2004-40503:PRL}. Cluster states based on photonic polarization qubits have been generated in bulk optical systems utilizing nonlinear optics~\cite{Walther:2005-169:NAT,Chen:2007-120503:PRL,Prevedel:2007-65:NAT,Tokunaga:2008-210501:PRL,Wunderlich:2011-33033:NJP,Ciampini:2016-e16064:LSA} or periodically driven quantum dots~\cite{Schwartz:2016-434:SCI} to achieve the required entanglement between multiple photons. The basic elements of quantum computation have been demonstrated with these polarization qubit states, including qubit rotation, two-qubit gates and small scale versions of algorithms, such as Grover's algorithm~\cite{Walther:2005-169:NAT}. Photonic cluster states can also be created using continuous variable quantum entanglement~\cite{Yokoyama:2013-982:NPHOT,Chen:2014-120505:PRL,Alexander:2016-32327:PRA}, where the qubits are encoded in the time dependent quadrature of the field. Furthermore, large cluster states have been demonstrated on-chip using frequency space encoding of the qubits \cite{Kues:2017-622:NAT}. Here we consider the generation of cluster states using a fully spatial encoding of each qubit, which is well suited for on-chip implementation.

%although we will focus our attention on discrete systems of spatially entangled photons.

Typically photonic cluster states are generated in bulk optical setups by passing a pulsed pump laser twice through a nonlinear crystal, generating a pair of photons by spontaneous parametric down-conversion (SPDC) on each pass giving a four photon polarization entangled state~\cite{Walther:2005-169:NAT}. Alternatively just two photons can be used, since by exploiting hyper-entanglement in spatial and polarization degrees of freedom four qubits can be encoded in the two photons~\cite{Chen:2007-120503:PRL,Wunderlich:2011-33033:NJP}. This approach has the advantage of producing higher photon count rates, while still producing nontrivial four qubit cluster states. Encoding multiple qubits in a single photon has the potential to significantly increase the size of cluster states that can be realized, and it has been shown that such states can be used for quantum algorithms~\cite{Chen:2007-120503:PRL}. However, it is important to note that having multiple qubits encoded into a single photon decreases the flexibility of the overall system to execute arbitrary algorithms, since pairs of qubits encoded into a single photon must always be measured at the same time.

So far the realization of spatially encoded cluster states has been largely restricted to bulk optical setups, but inevitably scalable cluster state generation will require full on-chip integration. Cluster states based on hyper-entanglement between polarization and spatial degrees of freedom have been demonstrated on-chip~\cite{Ciampini:2016-e16064:LSA}, but a more natural and convenient realization would be based on just the spatial degree of freedom, since processing orthogonal polarizations in the same waveguide requires highly specialized fabrication platforms. In principle existing silicon photonics approaches \cite{Qiang:2018-534:NPHOT} could generate spatially encoded cluster states through use of reconfigurable linear optics to tune the wavefunction created by on-chip photon sources. However, it is interesting to consider the design of nonlinear sources which directly 
generate cluster states from a nonlinear interaction, with no linear optical post-processing step.
%, whereas fabricating multiple waveguides supporting the same single polarization is comparatively straight-forward.

Here, we describe a method for the generation of cluster states within a nonlinear photonic chip with no reconfigurable elements, where the state is fully encoded in the spatial properties of the photons. This method allows switching between different cluster states all optically, without a need for complex reconfigurable components to be integrated on-chip. The paper is organized as follows. In Sec.~\ref{sec:nonlinear} we introduce our general method for the generation of spatially entangled photon states in arrays of coupled nonlinear waveguides. Then, in Sec.~\ref{sec:domain} we specify how the domain poling patterns in the waveguides can be designed to produce cluster states. In Sec.~\ref{sec:algotithms} we demonstrate how simple computations could be carried out on the generated cluster states. Finally we present conclusions and outlook in Sec.~\ref{sec:conclusions}.

\section{Nonlinear spatial entanglement generation} \label{sec:nonlinear}

We consider a photonic chip with second-order nonlinearity so that pairs of photons can be generated via type I spontaneous parametric down-conversion (SPDC)~\cite{Solntsev:2012-23601:PRL,Solntsev:2017-19:RPH}. It has been shown that SPDC in arrays of coupled waveguides provides a stable source of highly entangled two-photon states~\cite{Solntsev:2014-31007:PRX}, and that $\chi^{(2)}$ poling in the array can be engineered to produce tailored two-photon quantum states~\cite{Titchener:2015-33819:PRA}, thus it is a natural platform to consider for the generation of cluster states. Similarly to Refs.~\cite{Chen:2007-120503:PRL,Wunderlich:2011-33033:NJP} we propose to use two-photon states to encode four-qubit cluster states. However, instead of exploiting hyper-entanglement between polarization and spatial modes, we will use spatially distributed entanglement across an array.
%of waveguides.

\begin{figure}[t]
	%\begin{figure}[htbp]
	\centering
	\includegraphics[width=\linewidth]{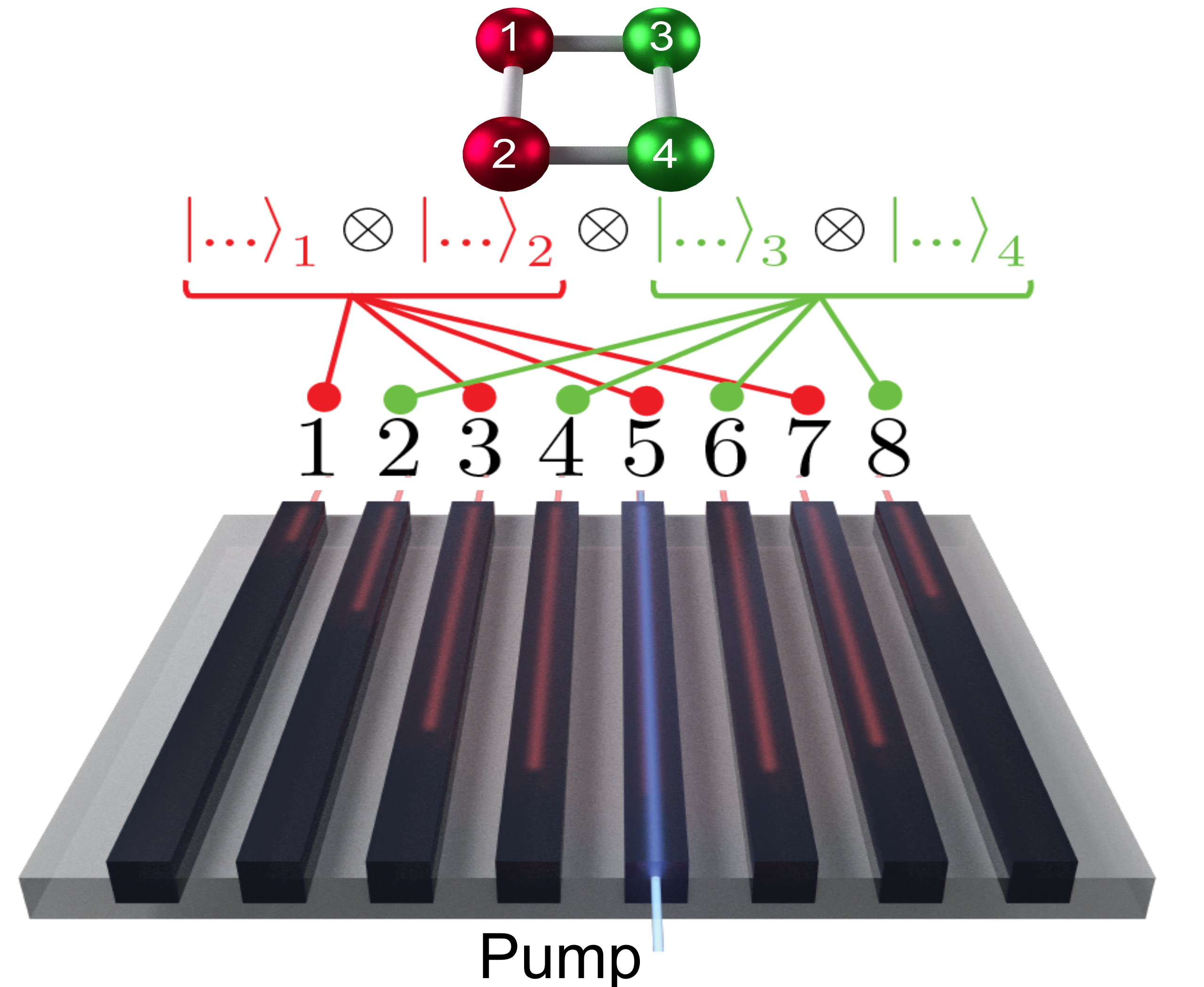}
	\caption{Conceptual diagram of a cluster-state encoded in an array of 8 waveguides. The array is driven by a pump laser (blue) producing a pair of entangled photons (red). The encoding from the two-photon eight-waveguide state to a four qubit cluster state is illustrated, whereby the signal (idler) photon in the odd (even) number waveguides encodes qubits $1$ and $2$ ($3$ and $4$) of the four qubit cluster state. }
	\label{fig:1}
\end{figure}

We demonstrate a potential for cluster state generation in an array of eight nonlinear waveguides, where specially tailored $\chi^{(2)}$ poling allows the production of specific two-photon states via SPDC.
We consider the regime where the signal and idler photons are indistinguishable spectrally, while their state is engineered such that observing one photon in odd numbered waveguides guarantees the other is in an even numbered waveguide.
Thus the system consists of two photons, each with four states available to it, giving a total of 16 distinct two-photon states.
As shown in Fig.~\ref{fig:1}, each of these two-photon states can be mapped to a different 4-qubit state, by encoding two computational qubits into the state of each photon. For example the physical two-photon state $\ket{1}_{\text{odd}}\ket{2}_{\text{even}}$, with one photon in waveguide 1 and the other in waveguide 2, would correspond to the 4-qubit state $\ket{0}_1\ket{0}_2\ket{0}_3\ket{0}_4$. Here the state of computational qubits $\ket{...}_1$ and $\ket{...}_2$ is defined by the physical state of the down-converted photon in the four odd numbered waveguides ($\ket{...}_{\text{odd}}$), and qubits $\ket{...}_3$ and $\ket{...}_4$ are defined by the state of the other down-converted photon in the four even numbered waveguides. Thus four qubit cluster states can be generated in the 8 waveguide system when the two-photon spatial wavefunction is shaped accordingly.

%% States encoded
%Using the encoding system in Fig.~\ref{fig:1}, we design an array of coupled waveguide that can produce both a star cluster state, and a box cluster state, as shown in Fig \ref{fig:2}. Pumping the first waveguide in the array will generate a two-photon pair in the star cluster state via SPDC, while pumping the last waveguide produces a two-photon pair in the box cluster state. This is possible since both waveguides are given different, specially tailored, inhomogeneous $\chi^{(2)}$ poling patterns. We consider the regime where the pump laser has no significant evanescent coupling to neighboring waveguides but the down-converted photons couple strongly and spread out across the whole array. Therefore when pumping different waveguides the pump laser will interact with different nonlinear poling patterns, producing different output down-converted two-photon states.

\section{Domain poling and cluster states} \label{sec:domain}

In order to shape the wavefunction in the array we
propose to use tailored domain poling patterns to allow control of the local effective nonlinearity along the pumped waveguide. This effectively defines the local phase of SPDC photon-pair generation at different points along the waveguide. In combination with the continuous photon-pair coupling to neighboring waveguides, this allows tailoring of the output spatial wavefunction. Similar methods have allowed wavefunction engineering in specially poled bulk nonlinear crystals~\cite{Yu:2008-233601:PRL,Qin:2008-63902:PRL,Leng:2011-429:NCOM} and in arrays of up to four coupled waveguides~\cite{Titchener:2015-33819:PRA,Lenzini:2018-17143:LSA}. Here we show how such control of the wavefunction can be achieved in eight
 coupled waveguides using only a single size of inverted $\chi^{(2)}$ domains, making fabrication of the structures feasible to implement with existing technology.
Using special domain poling patterns to control the two-photon wavefunction can be preferable to adjusting linear properties of the chip such as the intra-waveguide coupling rate.
 Furthermore, each waveguide can be given a different nonlinear poling pattern, allowing the chip to be quickly reconfigured to produce different cluster states simply by driving different waveguides in the array. This avoids the need to integrate complex thermal or electro-optic phase shifters onto the chip to reconfigure the wavefunction. Thus, inhomogeneous waveguide poling provides a straight-forward approach to generating and reconfiguring different photonic wavefunctions on-chip.

\begin{figure}[t]
	\centering
	\includegraphics[width=\linewidth]{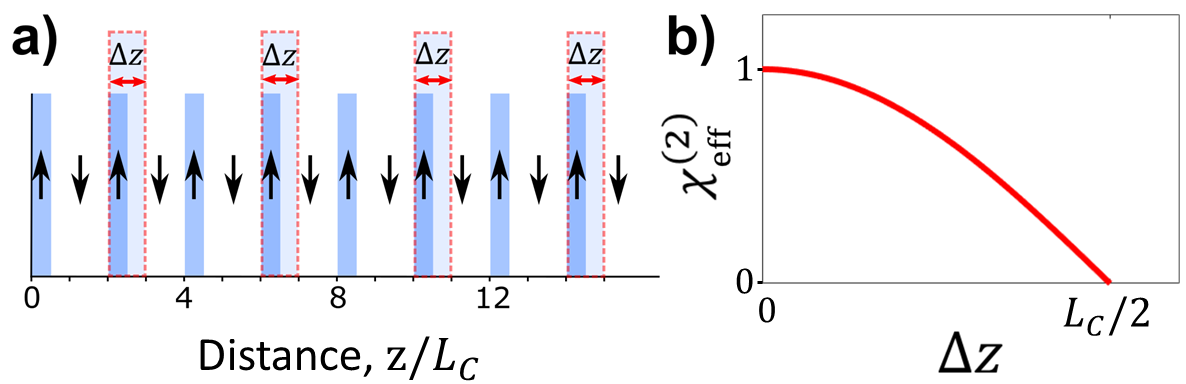}
	\caption{a) Diagram of the poling technique. Every second `up' domain can be translated by $\Delta z$ to alter the local effective nonlinearity. b) The local effective nonlinearity (normalized to unity)   vs. the domain translation ($\Delta z$), showing full control with a translation of half a coherence length ($L_C/2$).}
	\label{fig:2}
\end{figure}

% Poling pattern:
To this end we develop a class of nonlinear poling patterns that give precise control over the local effective nonlinear coefficient of each waveguide. Particularly we focus on designing patterns that would be easy to fabricate, thus avoiding varied domain sizes such as in Refs.~\cite{Qin:2008-63902:PRL,Titchener:2015-33819:PRA}. This is achieved by superimposing two fourth-order periodic poling patterns [Fig.~\ref{fig:2}(a)]. Fourth-order patterns have period equal to four times the decoherence length of the SPDC process ($4\, L_C$), and in this case we consider patterns where the `up' domain length is $0.5\,L_C$, and the remaining $3.5\,L_C$ is poled down. Two of these patterns are then superimposed, as in Fig.~\ref{fig:2}(a), to make a second-order phase-matched poling pattern. The displacement between the two fourth-order patterns, $\Delta z$, determines the phase difference between the two-photon wavefunction generated from each poling pattern. Accordingly, the local effective nonlinearity of the poling structure can be controlled by varying this displacement as shown in Fig.~\ref{fig:2}(b). Translating the whole structure (with respect to other sections of poling on the waveguide) changes the overall phase of the wavefunction generated from that section of poling. Therefore the displacement of overlapping fourth-order patterns with respect to one another ($\Delta z$) controls the magnitude of the effective nonlinearity, while translating the whole structure controls the phase.

% Practicality of poling pattern
A key consideration for ferroelectric domain poling is that the size of the domains to be inverted should be the same for the entire chip, and the inverted domains should not be too close together. This is because the growth of ferroelectric domains is a complex process, and fabrication parameters such as electrode size must be determined empirically to produce the required domain size~\cite{Houe:1995-1747:JPD}. Thus it is difficult to fabricate domains with a range of different sizes on the chip, and domains spaced too closely together can interact, or fuse together during the inversion process, producing unpredictable results. Since the general poling pattern we propose requires only `up' domains of a constant size, and these domains are never closer to one-another than $L_C/2$, it should be straightforward to fabricate using typical electric field poling with lithographically defined electrode masks.

\begin{figure}[t]
	\centering
	\includegraphics[width=\linewidth]{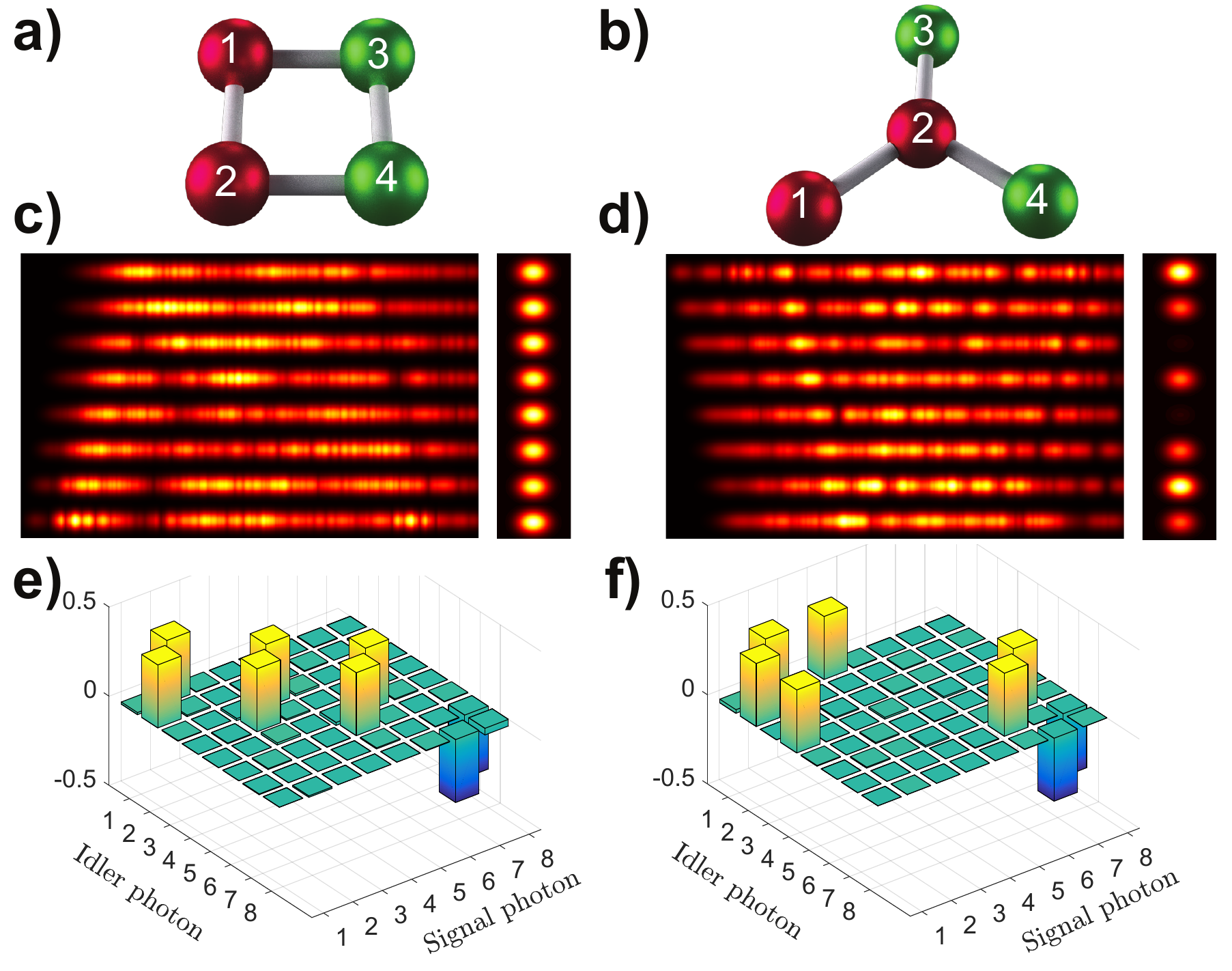}
	\caption{
 a) and b) Structure of the box and star cluster states, where red (green) spheres represent qubits physically encoded in the signal (idler) photon.
 c) and d) down-converted field intensity during the generation of the box and star cluster states respectively, using special waveguide array poling described in Fig.~\ref{fig:2}. e) and f) show the corresponding output two-photon wavefunctions produced when pumping waveguides $8$ and $1$ of the array. e) has a fidelity of $99.8\%$ to the perfect box cluster state and f) has a fidelity of $99.9\%$ to a perfect star cluster state.
}
	
	\label{fig:3}
\end{figure}

% Use of poling to tailor wavefunction.
In order to create tailored wavefunctions using this poling technique we divide each waveguide in the nonlinear waveguide array into 34 different sections, and allow each section to have a different poling pattern of the form shown in Fig.~\ref{fig:2}(a), and thus a different effective nonlinearity. Through algorithmic optimization of the effective nonlinearity in each section, we can design tailored poling structures to produce, via SPDC, a desired two-photon state at the output of the array. For a practical source of cluster states we design an eight waveguide device that produces the box or star cluster states [Figs.~\ref{fig:3}(a) and (b)] when waveguides number 8 or 1 respectively are driven by the pump laser. The down-converted photon intensity in the device is simulated in Figs.~\ref{fig:3}(c) and d) for production of the box and star cluster states respectively. The full output wavefunctions produced from the poling structures are presented in Figs.~\ref{fig:3}(e) and (f), with fidelity to the ideal box and star cluster states of $99.8\%$ and $99.9\%$, respectively.

\section{cluster state algorithms} \label{sec:algotithms}

% States encoded

%Using the encoding system in Fig.~\ref{fig:1}, we design an array of coupled waveguide that can produce both a star cluster state, and a box cluster state, as shown in Fig \ref{fig:2}. Pumping the first waveguide in the array will generate a two-photon pair in the star cluster state via SPDC, while pumping the last waveguide produces a two-photon pair in the box cluster state. This is possible since both waveguides are given different, specially tailored, inhomogeneous $\chi^{(2)}$ poling patterns. We consider the regime where the pump laser has no significant evanescent coupling to neighboring waveguides but the down-converted photons couple strongly and spread out across the whole array. Therefore when pumping different waveguides the pump laser will interact with different nonlinear poling patterns, producing different output down-converted two-photon states.

% Change of measurement basis.
Once the photons are created, cluster state computation algorithms proceed by sequentially measuring different qubits in the state. The measurement basis used to measure qubit number $i$ is denoted $B_i(\alpha)$, with basis states $\ket{\psi(\pm \alpha)}_i = \left(\ket{0}_i \pm e^{i\alpha}\ket{1}_i \right)/\sqrt{2}$, where the value of $\alpha$ is adjusted for each qubit measurement, depending on the algorithm being implemented~\cite{Raussendorf:2003-22312:PRA}. In the context of a waveguide array this measurement basis is non-trivial to implement, because it requires spatial transformations on the output waveguides to rotate to the $B(\alpha)$ measurement basis, regardless of the value of $\alpha$. However we observe that for many simple operations, such as propagating a state through a circuit, or performing a CNOT gate, only measurements in the bases $B(0)$ and $B(\pi)$ are required~\cite{Raussendorf:2003-22312:PRA,Nielsen:2006-147:RMAP}. Under a Hadamard transformation, these measurement bases are mapped to direct measurements in the waveguide mode basis, i.e. $\boldsymbol{\hat{H}} \left(\ket{0}_i + e^{i\pi}\ket{1}_i\right)/\sqrt{2}=\ket{1}_i$ and $\boldsymbol{\hat{H}} \left(\ket{0}_i - e^{i\pi}\ket{1}_i\right)/\sqrt{2}=\ket{0}_i$, so there is no need to perform any linear transformation before measurement. Thus in order to implement spatially encoded cluster state algorithms more efficiently in our proposed encoding scheme, a Hadamard transformation should be applied to all the measurement bases used for the algorithm, and also, to preserve the form of the algorithm, Hadamard transformations should be applied to each qubit in the cluster states itself. Thus the cluster states we designed above are the typical cluster states, but with a Hadamard transformation applied to each qubit in the state.

%\section{cluster state use and operation}

After Hadamard transformations are applied to each qubit of the star cluster state the resulting wavefunction is,
\begin{equation}
C_4^{star} =
\ket{0}_1 \ket{0}_2 \ket{0}_3 \ket{+}_4
\boldsymbol{+}
\ket{1}_1 \ket{1}_2 \ket{1}_3 \ket{-}_4,
\end{equation}
where $\ket{\pm}_i = \ket{0}_i \pm \ket{1}_i$. The corresponding spatial two-photon wavefunction of this state is shown in Fig.~\ref{fig:3}(f). Such a state could be used to implement a CNOT gate~\cite{Raussendorf:2003-22312:PRA}, provided the measurement basis is the Hadamard transformation of the typical basis.
%
% Box cluster state use and operation
%The four-qubit cluster state is formed starting with four qubits, all the in $\ket{+} = \ket{0}+\ket{1}$ state, then applying entangling controlled phase gates between the qubits [Fig.~\ref{fig:2}(a)]. The application of a controlled phase gate between two qubits is represented by a line joining those qubits in the diagram, in this case forming a box shape.
%
Similarly we define the box cluster state as,
\begin{multline}
C_4^{box} =
\ket{0}_1 \ket{+}_2  \ket{0}_3 \ket{+}_4  \boldsymbol{+}
\ket{1}_1 \ket{-}_2  \ket{0}_3 \ket{-}_4 \boldsymbol{+} \\
\ket{0}_1 \ket{-}_2  \ket{1}_3 \ket{-}_4 \boldsymbol{+}
\ket{1}_1 \ket{+}_2  \ket{1}_3 \ket{+}_4,
\end{multline}
and the representation of this state as a two-photon spatial wavefunction is shown in Fig.~\ref{fig:3}(e).

	% operation of Gorver's search
The box cluster state can be used for an implementation of Grover's search algorithm~\cite{Grover:1997-325:PRL, Walther:2005-169:NAT, Prevedel:2007-65:NAT,Chen:2007-120503:PRL}. For the simple case of a two qubit database this search consists of two steps. First a two bit state is prepared in the $\ket{+}\ket{+}$ state, and a two-bit string to be recovered (e.g. `$01$') is encoded into the  state by inverting the sign of the corresponding wavefunction element (e.g. $\ket{0}\ket{1}$). In the next step the amplitude of the quantum state representing this encoded string is amplified by inverting the entire state about the mean. For the two qubit case the answer is produced in a single iteration.

%\section{Implementation on box cluster}

\begin{figure}[t]
	\centering
	\includegraphics[width=\linewidth]{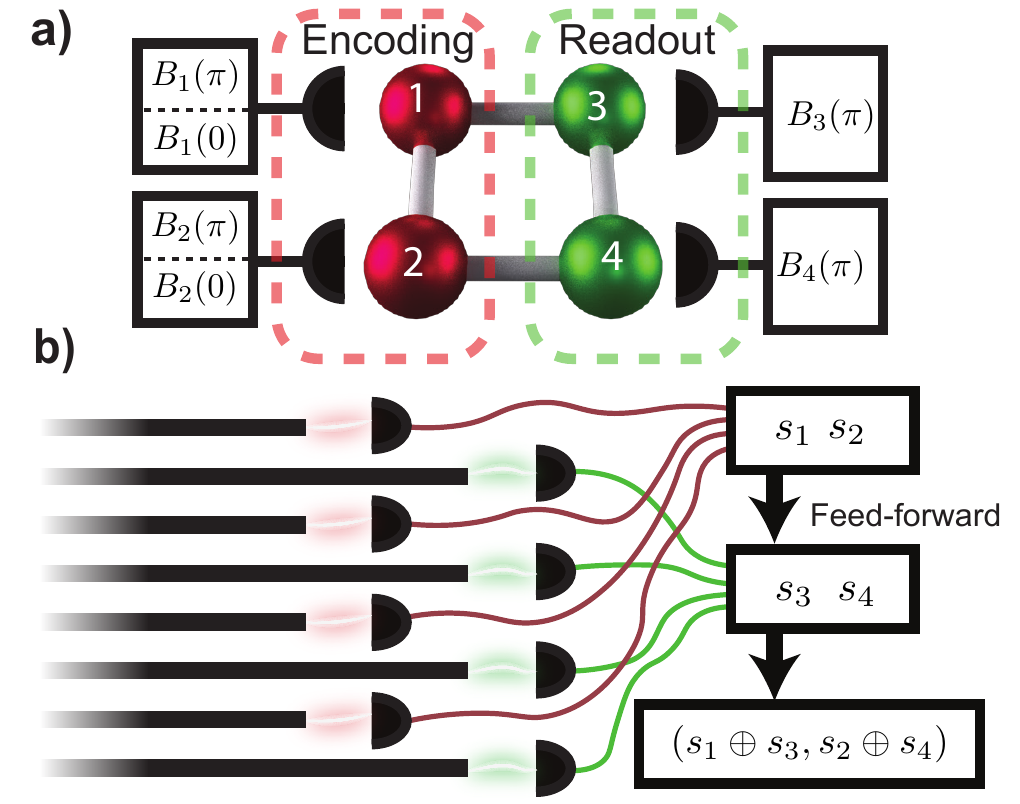}
	\caption{Operation of Grover's search algorithm. a) Cluster state diagram of the implementation of Gorver's search. b) Implementation in the 8-waveguide system. Measurements are made directly in the waveguide output mode basis.}
	\label{fig:4}
\end{figure}

	The measurements required to implement this algorithm using the box cluster state are shown in Fig.~\ref{fig:4}(a), with our implementation in the eight waveguide spatial encoding shown in Fig.~\ref{fig:4}(b).
    As discussed above, for spatially encoded cluster states we propose to use a measurement basis that is the Hadamard transform of the usual measurement basis. We now define the basis explicitly as $B^{\hat{H}}_i(\alpha)$ with basis states $\ket{\psi(\pm \alpha)}_i = \left(\ket{+}_i \pm e^{i\alpha}\ket{-}_i \right)/\sqrt{2}$, where detection of one of the two basis states is interpreted as a logical $0$ or $1$ respectively. To implement the Grover's search algorithm in this spatial encoding, at first the qubits $1$ and $2$ are measured, which physically involves detecting which odd number waveguide the signal photon is in. The choice of measurement basis determines the bit string that is marked for recovery.
 Measuring qubit $i\in\{1,2\}$ in the basis $B_i^{\hat{H}}(\pi)$ (or $B_i^{\hat{H}}(0)$) will encode a logical $0$ (or $1$) into the $i_{\text{th}}$ element of the bit string to be recovered. If both measurement results, $s_1$ and $s_2$, are $0$ the initial encoding of the two-bit string was successful, otherwise unsuccessful encoding can be compensated for by feeding forward the measurement results and using them to rotate the measurement bases for qubits $3$ and $4$.
To recover the encoded bit string  via Grover's search algorithm the remaining two qubits, $3$ and $4$, are measured in the basis $B^{\hat{H}}_{i}(\pi)$, physically achieved by detecting the idler photon in one of the even numbered waveguides. 	In place of rotating the measurement basis of qubits $3$ and $4$ post-processing of results can instead be used, reinterpreting final result as $\left(s_1\oplus s_3, s_2 \oplus s_4\right)$. This recovers the marked bit string with certainty.

 For example, if we choose to encode the bit string `$01$' into the two-photon state, for recovery we would use the measurement bases $B_1^{\hat{H}}(\pi)$ and $B_2^{\hat{H}}(0)$. Detection of the signal photon in waveguide $1$ corresponds to the state $\ket{0}_1\ket{0}_2$, which in these measurement bases is interpreted as measurement results $s_1=1$ and $s_2 = 0$. Then due to the cluster state structure [Fig.~\ref{fig:3}(e)] the idler photon will be detected in waveguide $2$, corresponding to state $\ket{0}_3\ket{0}_4$, which is interpreted with the required basis $B_3^{\hat{H}}(\pi)$ $B_4^{\hat{H}}(\pi)$ to give results $s_3 = 1$, $s_4 = 1$. Finally, the recovered bit string is $\left(s_1\oplus s_3, s_2 \oplus s_4\right) = (0, 1)$, exactly the bit string that was encoded by the measurement of qubits $1$ and $2$.

\section{Conclusions}
\label{sec:conclusions}

In conclusion, we have shown how to design a nonlinear photonic chip to generate and optically switch between different 4-qubit cluster states. This is achieved using a nonlinear waveguide array with specially tailored poling patterns, which are optimized to be easy to fabricate with typical electric field poling methods.
Importantly this can provide a stable integrated source of cluster states, with potential to scale to larger states by increasing the number of waveguides in the array.
We also propose a change of measurement basis to implement the cluster state algorithms with spatially encoded photonic qubits.

%\paragraph{Funding.}
\section*{Acknowledgments}
We acknowledge funding from the Australian Research Council (ARC) projects DP160100619, DP190100277, and DE180100070. The authors thank Mirko Lobino for valuable discussions.

% Bibliography
\bibliography{db_Cluster_states_paper_v2}
%\newpage
%\bibliographyfullrefs{Cluster_states}

%\ifthenelse{\equal{\journalref}{ol}}{%
%\clearpage
%\bibliographyfullrefs{db_sparse_tomography_v3}
%}{}

\end{document}